\documentclass[a4paper,11pt]{article}

\usepackage{pos}
\usepackage{lipsum} 
\usepackage{multirow}
 \usepackage{caption}
   \usepackage{subcaption}
\title{IceCube population constraints on neutrino emission by Fermi-LAT detected active galactic nuclei}

\ShortTitle{IceCube population constraints on neutrino emission by Fermi-LAT detected active galactic nuclei}

\author{The IceCube Collaboration \\{\normalsize \normalfont(a complete list of authors can be found at the end of the proceedings)}\\}

\emailAdd{sahori@wisc.edu}
\emailAdd{abhishek.a.desai@nasa.gov}
\emailAdd{justin.vandenbroucke@wisc.edu}

\abstract{

Gamma-ray-bright active galactic nuclei (AGN) have been one of the most promising source classes of high-energy astrophysical neutrinos detected by IceCube. The first evidence of an IceCube point source was a blazar detected by the Fermi Large Area Telescope (LAT), TXS ~0506+056. Previous analyses have ruled out GeV-bright blazars as the predominant contributor to the high-energy astrophysical neutrino flux under simple correlation assumptions about the relationship between the fluxes of gamma rays and neutrinos. We present results from a more general and more sensitive search for correlation between neutrinos and GeV-selected AGN using improvements in the IceCube statistical methods and 13 years of data.  We detect no correlation and set stringent constraints on neutrino emission by populations of GeV-detected AGN.  These include constraints on the neutrino emission from subclasses of GeV-detected AGN, including BL Lacs, flat-spectrum radio quasars (FSRQ) and non-blazar AGN, using stacking analyses testing a variety of hypothesized relationships between neutrino and gamma-ray flux.  We also present results from an analysis that is sensitive to a wider range of relationships between the gamma-ray and neutrino signal.

\vspace{4mm}

{\bfseries Corresponding authors:}
Sam Hori$^{1*}$, 
Abhishek Desai$^{2}$, 
Justin Vandenbroucke$^{1}$
\\
{$^{1}$ \itshape Dept. of Physics and Wisconsin IceCube Particle Astrophysics Center, University of Wisconsin{\textemdash}Madison, Madison, WI 53706, USA}\\
{$^{2}$ \itshape NASA Postdoctoral Program Fellow, NASA Goddard Space Flight Center, Greenbelt, MD, 20771, USA}\\
$^*$ Presenter
}

\ConferenceLogo{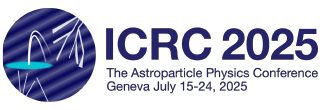}

\FullConference{39th International Cosmic Ray Conference (ICRC2025)\\
 15–24 July 2025\\
Geneva, Switzerland\\}

\begin{document}

\maketitle

\section{Background}\label{sec_background}

The IceCube Neutrino Observatory (IceCube) has identified a diffuse flux of astrophysical neutrinos \cite{IceCube:2013a} as well as the first sources of high-energy neutrinos. The first known IceCube-detected neutrino sources are heterogeneous; the BL Lac TXS 0506+056 \cite{Icecube_TXS_Alert_paper, Icecube_TXS_flare_paper}, a type 2 Seyfert NGC 1068 \cite{icecube_NGC1068Result}, and the Galactic Plane of the Milky Way \cite{IceCube_GalacticPlane}. The two detected point sources are both active galactic nuclei (AGN) but have very different properties. TXS 0506+056 is bright in gamma rays and was detected as a time-dependent source while NGC 1068 is not a blazar and the neutrino emission is not known to be time-dependent. Both NGC 1068 and TXS 0506+056 are sources of GeV gamma rays \cite{Fermi_4FGL_DR4}, but the gamma rays from NGC 1068 originate from a starburst component \cite{Eichmann_2022} rather than the core where the neutrinos are theorized to originate. Understanding the relationship between gamma rays and high-energy neutrinos in AGN is an important step in understanding AGN physics and the population which contributes to the IceCube diffuse flux.

Many models predict a correlation between high-energy gamma rays and neutrinos \cite{Yuan_2020}. Previous studies have constrained the neutrino flux from gamma-ray bright AGN under simple assumptions about the connection between gamma-rays and neutrinos \cite{IceCube_2LAC, IceCube_3FHL, IceCube_1FLE}. 
A larger neutrino dataset modeled with better reconstruction techniques have improved IceCube's sensitivity which combined with an increased Fermi-LAT livetime enhances the ability to probe the connection between gamma-ray and neutrino production in AGN.

\section{Data Samples} \label{sec_data}

\subsection{IceCube Dataset}
IceCube is a cubic-kilometer water Cherenkov neutrino detector located at the South Pole which uses the glacial ice as a detection medium and is optimized to detect TeV-PeV neutrinos \cite{Icecube_instrumentation}. 

IceCube classifies detected events as tracks and cascades. Tracks are primarily created by muons traveling through the detector. Tracks can be created by charged-current muon neutrino interactions or by muons originating from cosmic-ray interactions in the atmosphere (which provides the dominant source of background). Cascades are created by any process that deposits energy over a short distance, including charged-current $\nu_e$ or $\nu_\tau$ interactions or any neutral current interactions. In the southern sky, tracks result in an extremely background-dominated data sample due to a large flux of atmospheric muons created by cosmic-ray interactions in Earth's atmosphere. However, tracks have a significantly better angular resolution ($\lesssim 0.5$ degrees) than cascades ($\lesssim 10$ degrees), so the preferred topology for IceCube's searches for the astrophysical sources of neutrinos is tracks originating from the northern sky, where Earth is able to attenuate much of the cosmic-ray muon flux.  

This analysis uses 13 years of an IceCube northern track data selection (``Northern Tracks''), shared with a previous IceCube point source catalog search \cite {ICRC_proceeding_13_year}. A 10-year selection of the data provided the first statistically significant evidence that the Type 2 Seyfert NGC 1068 is a neutrino source \cite{icecube_NGC1068Result}. This selection improves over past IceCube data selections by incorporating a point spread function described by a kernel density estimator, rather than a Gaussian. The sample includes 991,499 events with a livetime of 13.05 years from June 2010 and November 2023 with sensitivity to sources with declinations between -3 and 81 degrees.

\begin{figure}[h!]
\centering
\begin{subfigure}{.48\textwidth}
  \includegraphics[width=1\linewidth]{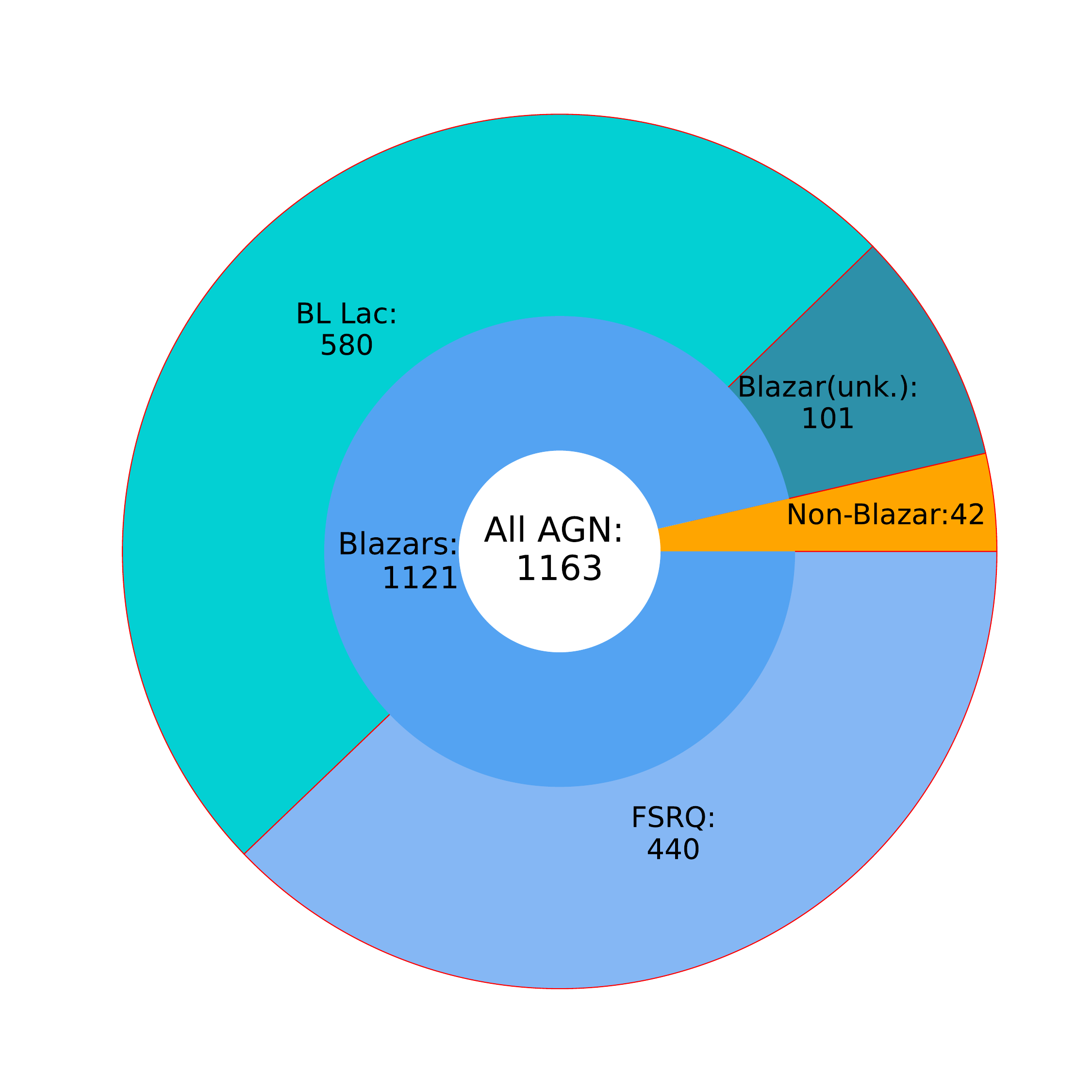}
\end{subfigure}%
\begin{subfigure}{.39\textwidth}
  \raisebox{.12\textwidth}{\includegraphics[width=1\linewidth]{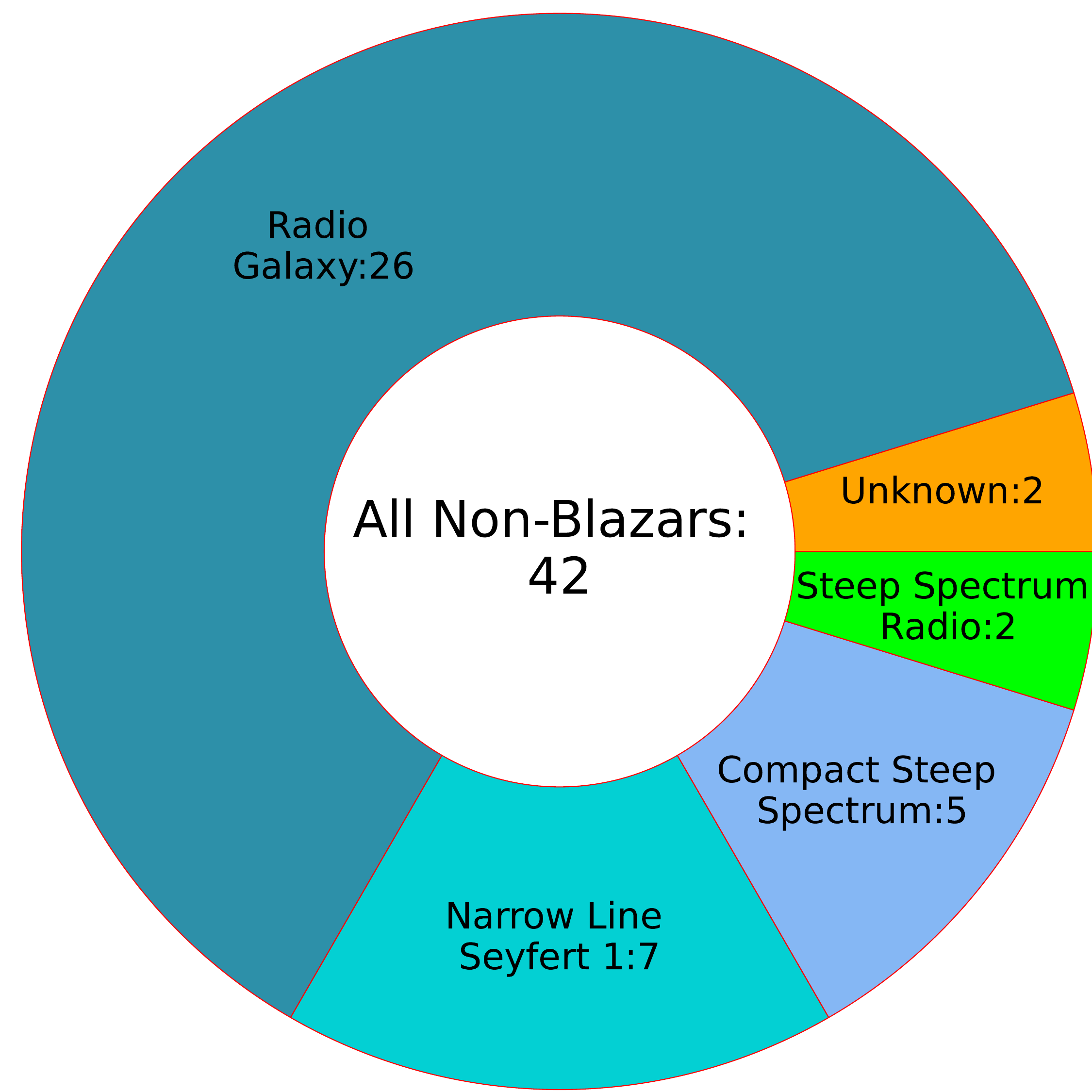}}
\end{subfigure}
\caption{Pie chart of AGN population after redshift and declination cuts are made. Left plot shows the break down into non-blazars and blazars, with blazars broken out into FSRQ, BL Lac and unknown class subgroups. The right plot breaks down non-blazars into more specific classifications}
\label{fig:pie_population}
\vspace{-.2in}

\end{figure}
\subsection{Fermi Catalog}

The analyzed AGN catalog is based on the Fermi 4LAC-DR3 \cite{Fermi_4LAC_DR3} and 4FGL-DR4 \cite{Fermi_4FGL_DR4} catalogs. Cuts are applied to ensure that each source is in both catalogs, has a redshift entry in the Fermi 4LAC-DR3, and has a declination between -3 and 81 degrees (due to the IceCube data selection). We use the spectrum and classification values provided in the 4FGL-DR4. 

After cuts, we are left with 1163 objects in the catalog. Figure \ref{fig:pie_population} shows the breakdown by class. This catalog has a more balanced number of FSRQ and BL Lac than the original catalogs because the redshift requirement removes more BL Lac objects than FSRQs. 42 objects are classified as various types of non-blazar AGN. These exclude sources where Fermi associates the gamma-ray flux from an AGN with a separate component, including NGC 1068 (where the gamma-rays are associated with a starburst component).

\section{Methods}\label{sec_methods}
\subsection{Stacking Test}
The first component of this analysis is a stacking test which looks for neutrino emission from a given source class under a model for the correlation between neutrinos and gamma rays. IceCube neutrino source analyses are done with a likelihood construction,
\begin{equation}
    \mathcal{L}(n_s,\gamma)=\prod_i^N \frac{n_s}{N}S_i(\alpha_i,\delta_i,\sigma_i,E_i,\gamma)+(1-\frac{n_s}{N})B_i(\delta_i,E_i)
\end{equation}
where $S_i$ and $B_i$ represent the probability that a source or background neutrino respectively would have the observed properties of the i-th neutrino. $n_s$ is the number of signal neutrinos and \textit{N} represents the total number of neutrinos in the dataset.  We test a power-law spectrum with $\frac{dN}{dE}\propto E^{-\gamma}$, where E is the energy of the neutrino. 

For the "Northern Tracks" dataset, the parametrization of $S_i$ and $B_i$ uses the kernel density estimator (KDE) approach described in \cite{ICRC_skyllh_proceeding}. For a stacking analysis, we follow a similar procedure to previous IceCube stacking analyses \cite{IceCube_hard_xray}. We weight each source by a neutrino expectation term $w^k$. In addition, we must account for a detector-dependent signal detection efficiency weighting term $R^k(\delta_i,\gamma)$. The $S_i$ term is then given by a sum over the \textit{M} sources in the sample,
\begin{equation}
    S_{i,stacking}=\frac{\sum_i^M w^k \cdot R^k(\delta_i,\gamma)\cdot S^k_{i,PS}(\alpha_i,\delta_i,\sigma_i,E_i,\gamma)}{\sum_i^M w^k\cdot R^k(\delta_i,\gamma)}.
\end{equation}
This stacking analysis tests a model in which the neutrino spectrum follows a power law spectrum with a shared index betweeen all sources. More complex models may make predictions of a spectral index which varies over the sources.

The p-value is computed using a log likeihood ratio test statistic
\begin{equation}
    TS=2\ln(\frac{max_{n_s,\gamma} \mathcal{L}(n_s,\gamma)}{\mathcal{L}(n_s=0)}).
\end{equation}
To create a Monte Carlo background TS distribution, the observed data is scrambled in right ascension repeatedly and the scrambled data is then evaluated. The p-value is then computed via comparison of the observed TS with the TS distribution generated under pseudo-trials of the null hypothesis. The analysis' response to a signal is obtained by injecting Monte Carlo simulated neutrinos on top of the scrambled background and evaluating the pseudo-trial.

In this analysis, we test several hypotheses about the correlations between neutrinos and gamma-ray emission. As a proxy for gamma-ray power, we test correlations between the neutrino luminosity and the gamma-ray rate between 1-100 GeV in the source frame. We separate the catalog into subsets of flat-spectrum radio quasars, BL Lac objects and non-blazar AGN. We weight based on the gamma-ray rate $R_\gamma^{\Gamma}$, the number of photons emitted between 1 and 100 GeV in the source frame. We tested models where the neutrino emission is either  proportional to the gamma-ray rate or the gamma-ray rate squared ($L_\nu\propto R_\gamma^{\Gamma}$ with $\Gamma=1,2$), where $L_\nu$ is the neutrino luminosity.  The $\Gamma=1$ case is motivated by scenarios in which the observed gamma rays are associated directly with pion decay (and the associated neutrino production) or where both the gamma-ray and neutrino production is correlated with the intrinsic jet power. The $\Gamma=2$ case is motivated by models where both the cosmic ray power and the target field scale with the jet power \cite{Tavecchio_spline_sheath}. This leaves us with a total of 6 stacking hypothesis, 3 classes (BL Lac, FSRQ, non-blazar AGN) with 2 correlation hypothesis ($\Gamma=1$, $\Gamma=2$) each. The significances for each of these tests is trials-corrected through comparison by comparing the most significant p-value for each class and comparing to the distribution of this statistic obtained in set of background joint Monte Carlo pseudo-trials.

\subsection{Binomial Test}
The stacking analysis is model-dependent. Models where neutrino and gamma-ray emission do not directly correlate or where only a subset of AGN are neutrino sources, the stacking test may miss a bright signal. This analysis probes this model-independent space with a binomial test. For this analysis we compute a point source p-value for every source in our catalog. We then order these p-values by magnitude and define a binomial probability $\alpha$ as a function of the (fit) number of sources ($k$):
\begin{equation}
    \alpha(k)=\sum_{l=k}^M \binom{M}{l} p_k^l \cdot (1-p_k)^{1-l}.
\end{equation}

The test statistic (TS) is the maximized binomial probability value over k:
\begin{equation}
    TS_{binomial}=max_{k} \alpha(k).
\end{equation}

Monte Carlo distributions of background binomial TS values are created through the data scrambling method and a p-value is computed by comparison to pseudo-trials under the null hypothesis. 

\section{Results}\label{sec_results}
\subsection{Stacking Analysis}

\begin{table}[h!]
\centering
\begin{tabular}{ |c|c|c|c|c|c|} 
\hline
 AGN Subclass & $\Gamma$ & local p-value & $n_s$ & $\gamma$ & Trials-corrected p-values \\
 \hline

 \multirow{2}{*}{BL Lac} & 1 & 0.009 & 10.36 & 1.9 & \multirow{2}{*}{0.014 $(2.2\sigma)$}  \\ 
  & 2 & 0.052 & 5.73 & 2.0 & \\ 
   \hline

 \multirow{2}{*}{FSRQ} & 1 &  0.122 & 4.26 & 2.0 & \multirow{2}{*}{0.15 $(1.0\sigma)$} \\ 
  & 2 & 0.107 & 5.32 & 2.9 & \\ 
 \hline

 \multirow{2}{*}{Non-Blazar AGN} & 1 & 0.365 & 6.57 & 3.1 &\multirow{2}{*}{0.55}  \\ 
 & 2 & 0 & 0 &2.64 &\\ 
\hline

\end{tabular}

 \caption{Table of results for stacking analysis including the tested subclass, the index relating neutrino luminosity to the gamma-ray rate, the p-value, the best-fit number of source neutrinos and spectral index, and the p-value corrected for the two tested hypotheses.}
 \label{Table_stackingresults}

\end{table}

\begin{figure}[htbp]
\centering
\includegraphics[width=1\linewidth]{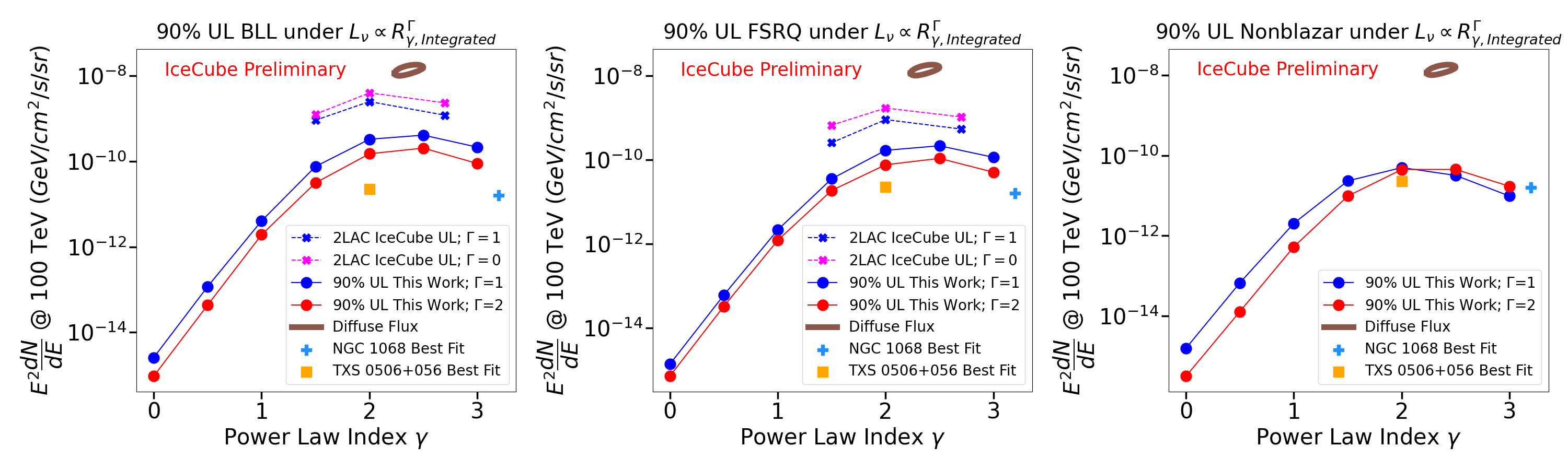}
\caption{
90\% upper limits for each source class under a model in which neutrino luminosity correlates with the number of gamma rays emitted between 1-100 GeV compared with the most analogous IceCube 2LAC-based results (Aartsen et al. 2017), the diffuse neutrino flux (Abbasi et al. 2022) and the best fit fluxes for TXS 0506+056 and NGC 1068 (Abbasi 2022). For BL Lacs, which are not tested as a single catalog in the previous 2LAC-based analysis, the upper limit is computed as the sum of the upper limits for low synchrotron peak BL Lac  and intermediate/high low synchrotron peak blazar classes. Neither upper limit result is corrected to account for unresolved sources in the source class.
}
\label{upper_limit_figure}
\vspace{-.2in}
\end{figure}

The results of the stacking analysis are contained in Table \ref{Table_stackingresults}. We do not find statistically significant results in any of the tested hypotheses.  We set upper limits to the tested models in Figure \ref{upper_limit_figure} and demonstrate that these limits are significantly stronger than the most analogous previous limits IceCube has placed on these population models \cite{IceCube_2LAC}, though the exact classes and hypotheses tested are not identical. While no correction has been made for the Fermi unresolved population in either upper limit, it is clear that the upper limits under a simple correlation between neutrinos and gamma-ray emission in AGN would only be able to provide a small fraction of the observed diffuse flux. These upper limits cannot rule out gamma-ray observed AGN as a source population under other correlation models or that gamma-ray AGN could be responsible for a fraction of the observed neutrino flux. The detection of neutrino flares from TXS 0506+056 indicates that some of the  diffuse flux must originate from gamma-ray bright AGN. 

\subsection{Binomial Test}
\begin{figure}[h!]
\begin{minipage}[c]{0.4\linewidth}
\centering
  \includegraphics[height=2in]{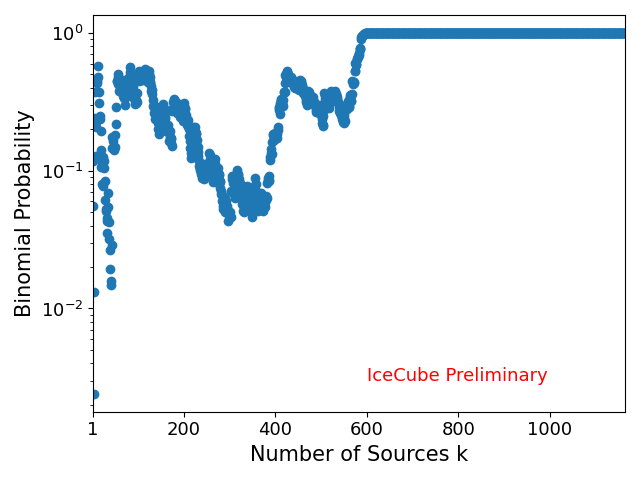}
  \caption{\centering
Binomial probability scanned over the number of sources.}
  \label{p_vs_k_binomial}

\end{minipage}
\hfill
\begin{minipage}[c]{0.4\linewidth}
\centering
  \includegraphics[height=2in]{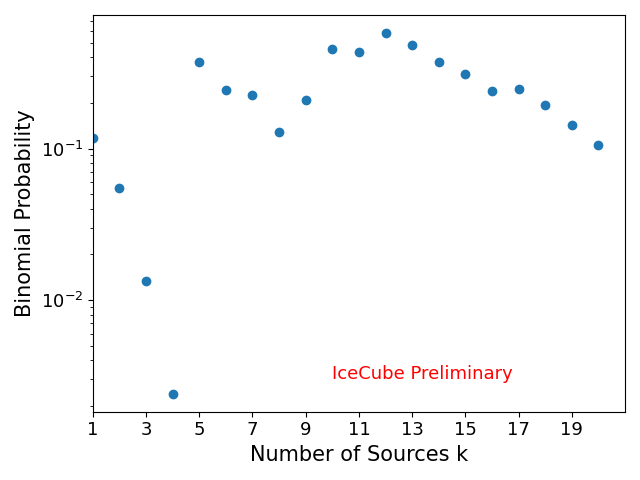}
  \caption{\centering
  Scan of binomial probability over the number of contributing sources zoomed in at low numbers of contributing sources near the observed best fit at k=4. }
  \label{p_vs_k_binomial_zoom}

\end{minipage}%
\end{figure}

The binomial test described in Section \ref{sec_methods} resulted in a p-value of 0.013 with a best-fit $k$ of 4. The variation of the binomial probability $\alpha(k)$ as a function of the fit number of contributing sources \textit{k} are shown in Figures \ref{p_vs_k_binomial},\ref{p_vs_k_binomial_zoom}. The 4 contributing sources were 4FGL J1427.0+2348 (PKS 1424+240), 4FGL J2322.7+3436 (TXS 2320+343), 4FGL J0509.4+0542 (TXS 0506+056) and 4FGL J1210.3+3928. NGC 1068 was not tested because the gamma-ray emission has been associated with the starburst component, so the Fermi AGN catalog excludes the source. PKS 1424+240 and TXS 0506+056 have been previously observed as some of the most significant IceCube subthreshold excesses \cite{icecube_NGC1068Result}. TXS 2320+343 appears to be driven by a high-energy neutrino in 2022, which was identified as a likely astrophysical neutrino by IceCube's realtime alert program and an alert was published over NASA's General Coordinate Network (GCN) \cite{TXS_2320+343_Alert}. 4FGL J1210.3+3928 is a BL Lac, but is notable for its proximity to NGC 4151, a nearby Seyfert galaxy. NGC 4151 has recently appeared as an interesting subthreshold excess in IceCube analyses \cite{IceCube_hard_xray,IceCube_xray_seyfert} examining the possibility of a Seyfert or X-ray bright AGN population of neutrino sources, which were motivated by evidence of NGC 1068 as a neutrino source. The proximity makes it difficult to distinguish the blazar from the Seyfert galaxy in either gamma-rays or neutrinos. Since NGC 4151 shares properties with NGC 1068 and any real astrophysical excess at the location of the blazar is most easily explained as source confusion with NGC 4151. 

\begin{figure}[h!]
\begin{minipage}[c]{0.5\linewidth}
\centering
  \includegraphics[height=2in]{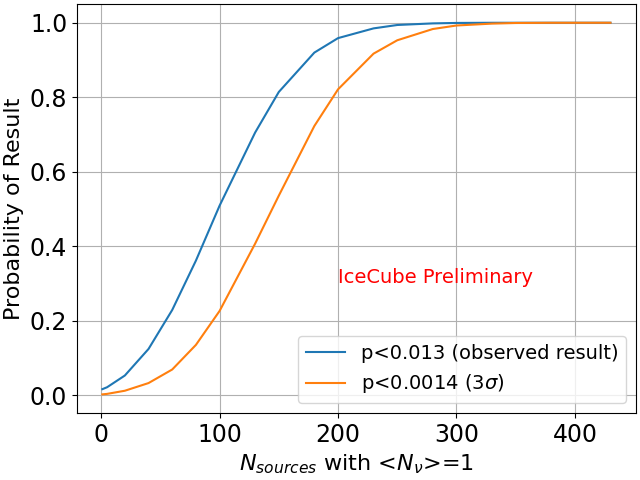}
      \captionsetup{width=.9\linewidth}

  \caption{
  \centering
Binomial test probability of obtaining a result more significant than either the observed result in the data or 3 sigma evidence under a model where a subset of the catalog are equal flux neutrino sources which contribute an average of one neutrino each as a function of the number of sources.}
  \label{binomial_sens1}
\end{minipage}
\begin{minipage}[c]{0.5\linewidth}
  \centering
  \includegraphics[height=2in]{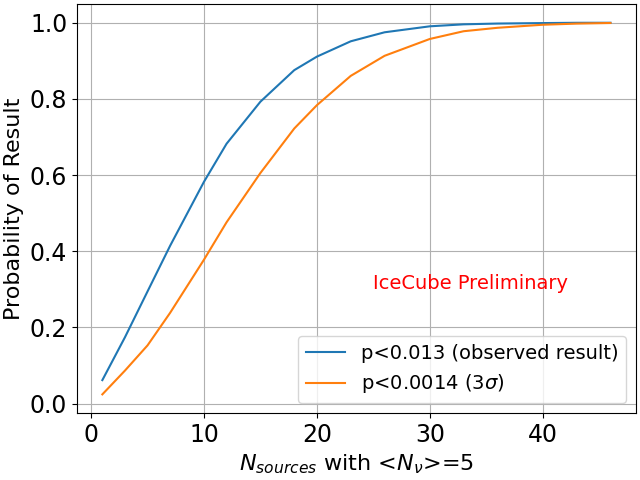}
    \captionsetup{width=.9\linewidth}
  \caption{\centering
  Binomial test probability of obtaining a result more significant than either the observed result in the data or 3 sigma evidence under a model where a subset of the catalog are equal flux neutrino sources which contribute an average of five neutrinos each as a function of the number of sources. }
  \label{binomial_sens5}
\end{minipage}
\end{figure}

Figures \ref{binomial_sens1},\ref{binomial_sens5} demonstrates the sensitivity to a nominal model in which a randomly chosen subset of the catalog each contribute a flux with a power-law spectrum with index of 2.0, with the flux normalization chosen to provide a fixed average number of neutrinos for each source (subject to Poisson uncertainty). The tested number of neutrinos are chosen so that objects with that flux would likely have been missed in previous catalog searches, but the population would contribute to the diffuse flux. This can be used to place constraints on that nominal model. This result constrains all models in which there are a large number of neutrino-emitting gamma-ray bright AGN, but the correlation between gamma-rays and neutrinos hasn't been directly explored in a stacking analysis.

\section{Conclusion}\label{sec_conclusion}
Progress in IceCube's methods, especially improvements in the parametrization of the point spread function, and livetime have allowed us to improve the upper limits on the neutrino flux contributed by gamma-ray bright AGN. 

Under simple models for the neutrino emission of gamma-ray bright AGN, IceCube data rejects the hypothesis that gamma-ray bright AGN are a primary source class of high-energy neutrinos. It remains unclear whether gamma-ray bright AGN (or a subclass of gamma-ray bright AGN) represent a separate source class than Seyferts or whether the evidence for TXS 0506+056 as a neutrino source implies a more subtle connection between neutrino emission across a broad group of AGN.

Under a model-independent test we obtain a statistically insignificant result and corresponding limits on the neutrino emission, but observe an interesting list of contributing sources consisting of two sources previously reported as non-significant excesses in time-independent IceCube gamma-ray selected catalog searches (PKS 1424+240 and TXS 0506+056), a source driven by a nearby high signalness neutrino (TXS 2320+343) and a BL Lac (4FGL J1210.3+3928). IceCube has previously found two pieces of over $3\sigma$ evidence for TXS 0506+056 as a time-dependent neutrino source, including a multi-messenger correlation with a high signalness neutrino.  4FGL J1210.3+3928 is close to NGC 4151, a Seyfert object not in the tested sample for this analysis, but which other searches have claimed as a major contributor to evidence that Seyferts may contribute a significant fraction of the IceCube flux.

\bibliographystyle{ICRC}
\bibliography{references}

\providecommand{\href}[2]{#2}\begingroup\raggedright\begin{thebibliography}{10}

\bibitem{IceCube:2013a}
{\bfseries IceCube} Collaboration, \href{http://dx.doi.org/10.1126/science.1242856}{{\em Science} {\bfseries 342} no.~6161, (2013) 1242856}.

\bibitem{Icecube_TXS_Alert_paper}
{\bfseries The IceCube Collaboration and Fermi-LAT and MAGIC and AGILE and ASAS-SN and HAWC and H.E.S.S. and INTEGRAL and Kanata and Kiso and Kapteyn and Liverpool Telescope and Subaru and Swift/NuSTAR and VERITAS and VLA/17B-403 teams} Collaboration, \href{http://dx.doi.org/10.1126/science.aat1378}{{\em Science} {\bfseries 361} no.~6398, (2018) eaat1378}.

\bibitem{Icecube_TXS_flare_paper}
{\bfseries IceCube} Collaboration, \href{http://dx.doi.org/10.1126/science.aat2890}{{\em Science} {\bfseries 361} no.~6398, (2018) 147--151}.

\bibitem{icecube_NGC1068Result}
{\bfseries IceCube} Collaboration, \href{http://dx.doi.org/10.1126/science.abg3395}{{\em Science} {\bfseries 378} no.~6619, (2022) 538--543}.

\bibitem{IceCube_GalacticPlane}
{\bfseries IceCube} Collaboration, \href{http://dx.doi.org/10.1126/science.adc9818}{{\em Science} {\bfseries 380} no.~6652, (2023) 1338--1343}.

\bibitem{Fermi_4FGL_DR4}
S.~e.~a. Abdollahi, \href{http://dx.doi.org/10.3847/1538-4365/ac6751}{{\em The Astrophysical Journal Supplement Series} {\bfseries 260} no.~2, (Jun, 2022) 53}.

\bibitem{Eichmann_2022}
B.~Eichmann, F.~Oikonomou, S.~Salvatore, R.-J. Dettmar, and J.~B. Tjus, \href{http://dx.doi.org/10.3847/1538-4357/ac9588}{{\em The Astrophysical Journal} {\bfseries 939} no.~1, (Nov, 2022) 43}.

\bibitem{Yuan_2020}
C.~Yuan, K.~Murase, and P.~Mészáros, \href{http://dx.doi.org/10.3847/1538-4357/ab65ea}{{\em The Astrophysical Journal} {\bfseries 890} no.~1, (Feb, 2020) 25}.

\bibitem{IceCube_2LAC}
{\bfseries IceCube} Collaboration, \href{http://dx.doi.org/10.3847/1538-4357/835/1/45}{{\em The Astrophysical Journal} {\bfseries 835} no.~1, (Jan, 2017) 45}.

\bibitem{IceCube_3FHL}
M.~Huber, \href{http://dx.doi.org/10.22323/1.358.0916}{``{Searches for steady neutrino emission from 3FHL blazars using eight years of IceCube data from the Northern hemisphere},''} in {\em Proceedings of 36th International Cosmic Ray Conference {\textemdash} PoS(ICRC2019)}, vol.~358, p.~916.
\newblock 2019.

\bibitem{IceCube_1FLE}
{\bfseries IceCube} Collaboration, \href{http://dx.doi.org/10.3847/1538-4357/ac8de4}{{\em The Astrophysical Journal} {\bfseries 938} no.~1, (Oct, 2022) 38}.

\bibitem{Icecube_instrumentation}
{\bfseries IceCube} Collaboration, \href{http://dx.doi.org/10.1088/1748-0221/12/03/P03012}{{\em Journal of Instrumentation} {\bfseries 12} no.~03, (Mar, 2017) P03012}.

\bibitem{ICRC_proceeding_13_year}
{\bfseries IceCube} Collaboration, \href{http://dx.doi.org/10.22323/1.444.1060}{{\em PoS} {\bfseries ICRC2023} (2023) 1060}.

\bibitem{Fermi_4LAC_DR3}
{\bfseries Fermi} Collaboration, \href{http://dx.doi.org/10.3847/1538-4365/ac9523}{{\em The Astrophysical Journal Supplement Series} {\bfseries 263} no.~2, (Nov, 2022) 24}.

\bibitem{ICRC_skyllh_proceeding}
{\bfseries IceCube} Collaboration, \href{http://dx.doi.org/10.22323/1.395.1073}{{\em PoS} {\bfseries ICRC2021} (2021) 1073}.

\bibitem{IceCube_hard_xray}
{\bfseries IceCube} Collaboration, \href{http://dx.doi.org/10.3847/1538-4357/ada94b}{{\em The Astrophysical Journal} {\bfseries 981} no.~2, (Mar, 2025) 131}.

\bibitem{Tavecchio_spline_sheath}
F.~{Tavecchio} and G.~{Ghisellini}, \href{http://dx.doi.org/10.1093/mnras/stv1023}{{\em MNRAS} {\bfseries 451} no.~2, (Aug., 2015) 1502--1510}.

\bibitem{TXS_2320+343_Alert}
{\bfseries IceCube} Collaboration, {\em GRB Coordinates Network} {\bfseries 33094} (Dec., 2022) 1.

\bibitem{IceCube_xray_seyfert}
{\bfseries IceCube} Collaboration, \href{http://dx.doi.org/10.48550/arXiv.2406.07601}{{\em arXiv e-prints} (June, 2024) arXiv:2406.07601}.

\end{thebibliography}\endgroup

%

\clearpage

\section*{Full Author List: IceCube Collaboration}

\scriptsize
\noindent
R. Abbasi$^{16}$,
M. Ackermann$^{63}$,
J. Adams$^{17}$,
S. K. Agarwalla$^{39,\: {\rm a}}$,
J. A. Aguilar$^{10}$,
M. Ahlers$^{21}$,
J.M. Alameddine$^{22}$,
S. Ali$^{35}$,
N. M. Amin$^{43}$,
K. Andeen$^{41}$
% ...
\\
\\
% ...
$^{10}$ Universit{\'e} Libre de Bruxelles, Science Faculty CP230, B-1050 Brussels, Belgium \\
% ...
$^{16}$ Department of Physics, Loyola University Chicago, Chicago, IL 60660, USA \\
$^{17}$ Dept. of Physics and Astronomy, University of Canterbury, Private Bag 4800, Christchurch, New Zealand \\
% ...
$^{21}$ Niels Bohr Institute, University of Copenhagen, DK-2100 Copenhagen, Denmark \\
$^{22}$ Dept. of Physics, TU Dortmund University, D-44221 Dortmund, Germany \\
% ...
$^{35}$ Dept. of Physics and Astronomy, University of Kansas, Lawrence, KS 66045, USA \\
% ...
$^{39}$ Dept. of Physics and Wisconsin IceCube Particle Astrophysics Center, University of Wisconsin{\textemdash}Madison, Madison, WI 53706, USA \\
% ...
$^{41}$ Department of Physics, Marquette University, Milwaukee, WI 53201, USA \\
% ...
$^{43}$ Bartol Research Institute and Dept. of Physics and Astronomy, University of Delaware, Newark, DE 19716, USA \\
% ...
$^{63}$ Deutsches Elektronen-Synchrotron DESY, Platanenallee 6, D-15738 Zeuthen, Germany \\
$^{\rm a}$ also at Institute of Physics, Sachivalaya Marg, Sainik School Post, Bhubaneswar 751005, India \\
% ...

\subsection*{Acknowledgments}

\noindent
The authors gratefully acknowledge the support from the following agencies and institutions:
USA {\textendash} U.S. National Science Foundation-Office of Polar Programs,
U.S. National Science Foundation-Physics Division,
U.S. National Science Foundation-EPSCoR,
U.S. National Science Foundation-Office of Advanced Cyberinfrastructure,
Wisconsin Alumni Research Foundation,
Center for High Throughput Computing (CHTC) at the University of Wisconsin{\textendash}Madison,
Open Science Grid (OSG),
Partnership to Advance Throughput Computing (PATh),
Advanced Cyberinfrastructure Coordination Ecosystem: Services {\&} Support (ACCESS),
Frontera and Ranch computing project at the Texas Advanced Computing Center,
U.S. Department of Energy-National Energy Research Scientific Computing Center,
Particle astrophysics research computing center at the University of Maryland,
Institute for Cyber-Enabled Research at Michigan State University,
Astroparticle physics computational facility at Marquette University,
NVIDIA Corporation,
and Google Cloud Platform;
Belgium {\textendash} Funds for Scientific Research (FRS-FNRS and FWO),
FWO Odysseus and Big Science programmes,
and Belgian Federal Science Policy Office (Belspo);
Germany {\textendash} Bundesministerium f{\"u}r Forschung, Technologie und Raumfahrt (BMFTR),
Deutsche Forschungsgemeinschaft (DFG),
Helmholtz Alliance for Astroparticle Physics (HAP),
Initiative and Networking Fund of the Helmholtz Association,
Deutsches Elektronen Synchrotron (DESY),
and High Performance Computing cluster of the RWTH Aachen;
Sweden {\textendash} Swedish Research Council,
Swedish Polar Research Secretariat,
Swedish National Infrastructure for Computing (SNIC),
and Knut and Alice Wallenberg Foundation;
European Union {\textendash} EGI Advanced Computing for research;
Australia {\textendash} Australian Research Council;
Canada {\textendash} Natural Sciences and Engineering Research Council of Canada,
Calcul Qu{\'e}bec, Compute Ontario, Canada Foundation for Innovation, WestGrid, and Digital Research Alliance of Canada;
Denmark {\textendash} Villum Fonden, Carlsberg Foundation, and European Commission;
New Zealand {\textendash} Marsden Fund;
Japan {\textendash} Japan Society for Promotion of Science (JSPS)
and Institute for Global Prominent Research (IGPR) of Chiba University;
Korea {\textendash} National Research Foundation of Korea (NRF);
Switzerland {\textendash} Swiss National Science Foundation (SNSF).

\end{document}